\documentstyle{article}

\title{\bf The system of four reggeized gluons
 and\\ the three-pomeron
 vertex in the high  colour limit}
\author{M. Braun \\
 Dep. High-Energy Physics, University of St. Petersburg,\\
198904 St.Petersburg, Russia}
\date{May 1997}
\def\beq{\begin{equation}}
\def\eeq{\end{equation}}

\def\Tr {{\mbox Tr}}
\def\ddf{D_{2\rightarrow 4}}
\def\dtf{D_{3\rightarrow 4}}
\def\dff{D_{4\rightarrow 4}}

\begin{document}
\maketitle
\medskip
\centerline{\bf Abstract.}

The amplitude for 4 interacting reggeized gluons is studied in the
high-colour limit. The leading order amplitude is explicitly shown to
reduce to a pair of reggeons, i.e to a single BFKL pomeron.  The
next-to-leading order  diffractive amplitude is found to split into a
double pomeron exchange and triple pomeron contribution. The obtained
three-pomeron vertex is different from the originally proposed in [2].

\vspace*{3 cm}
{\Large \bf SPbU-97-9}

 \newpage
\section{ Introduction.}
The system of four reggeized gluons has been studied in detail by
J.Bartels [1] and J.Bartels and M.Wuesthoff [2] for the realistic case
of the number of colours
$ N=3 $. The results are quite complicated because of the
interconnection between many different colour channels. The amplitudes
obey a system of coupled integral equations for which the inhomogeneous
term is provided by a part which is supposed to reduce to a pair of
reggeons (the "reggeizing piece") coupled to the non-reggeizing piece by
a transitional vertex. This vertex presents some nice features: it
is symmetric in all the gluons, infrared stable, conformal invariant
[3] and vanishes if any of the extrnal momenta goes to zero. However
its definition is rather dubious: in fact to obtain it in the form
presented in [2] it was assumed that a pair of reggeized gluons may
combine into a single reggeized gluon  not only in the adjoint
representation but also in other ones, which does not agree with the
corresponding BFKL equation.

In this note we study the same problem under the simplifying assumption
$ N\rightarrow\infty $. As we already stated in [4], in the leading
approximation, the amplitude for any number of reggeized gluons reduces
to the one for only a pair of them, i.e to a BFKL pomeron. In the
language of [1,2] this means that the non-reggeizing piece is down by
at least
$ 1/N $ as compared to the reggeizing piece for the leading colour
configuration (a cylinder in the picture in which the gluon is
represented by a
$ q\bar q $ line). We demonstrate this fact in some detail for the
triple discontinuity of the 4-gluon amplitude in Sec. 2.

As a result, a possibility opens up to study the complicated Bartels
system for different colour channel amplitudes by perturbation in
$ 1/N $ starting from the fully reggeizing piece in the leading
approximation. In this manner one arrives at a uniquely
defined transitional vertex which is different from [2], although it
preserves some of its good features. In particular, it is also infrared
stable and vanishes for any of the external momenta going to zero.
Apart from this, a new contribution arises which corresponds to a
direct coupling of two pomerons to a
$ q\bar q $ loop. It has an eikonal form for fixed transversal
dimension of the loop. This fact was also conjectured in [4]. Thus
the
$ N\rightarrow\infty $ approach allows to uniquely separate
contributions from the double pomeron exchange and the triple pomeron
coupling, which remain hidden in the transitional vertex of [2].

\section{The 4 gluon system at $N\rightarrow\infty$}
\subsection{The 2 gluon system}
The results for 2 gluons are well-known and are not sensitive for the
$ N\rightarrow\infty $ limit. We only list here the main formulas to
be used in the following. In our notations we try to follow [1,2].

In the lowest (zero) order approximation the 2 gluon amplitude
(actually its discontinuity Mellin transformed to the complex angular
momentum $ j $) is given by the $ q\bar q $ loop with the two gluons
attached to it in all possible ways (4 diagrams in all). Their colour
indeces $ a_{1} $ and $ a_{2} $ enter into the colour trace $
\Tr\{t_{a_{1}}t_{a_{2}}\}=(1/2)\delta_{a_{1}a_{2}} $, where $
t_{a} $ is the colour of the quark.. Separating this trace and the
coupling $ g^{2} $ we write the zero order contribution  as \beq
D_{20}(k_{1},a_{1},k_{2},a_{2})=(1/2)\delta_{a_{1}a_{2}}g^{2}
(f(0,k_{1}+k_{2})+f(k_{1}+k_{2},0)-f(k_{1},k_{2})-f(k_{2},k_{1}))
\eeq
Here
$ f(k_{1},k_{2}) $ is the contribution from the diagram with the gluon
1 attached to
$ q $ and the gluon 2 attached to
$ \bar q $. Assuming that the external particle is colourless (the
photon or "onium") we have
$ f(k_{1},k_{2})=f(k_{2},k_{1}) $. We define a vacuum colour state of
two gluons by its colour wave function
\beq
|0\rangle=(1/N)\delta_{a_{1}a_{2}}
\eeq
Then we find a projection of the
$ q\bar q $ loop onto this state from (1)
\beq
\langle D_{20}|0\rangle=\frac{N^{2}-1}{N}g^{2}(f(0,0)-f(k_{1},k_{2}))
\simeq g^{2}N(f(0,0)-f(k_{1},k_{2}))\equiv D_{20}(1)
\eeq
In the last notation, to be extensively used in the following, only the
number of the gluon whose momentum acts as one of the variables is
indicated, the other momentum determined from the conservation law
(e.g.
$ k_{2}=-k_{1} $ for the forward amplitude).

The full 2-gluon amplitude
$ D_{2} $ satisfies the vacuum channel BFKL equation
\beq
S_{20}D_{2}=D_{20}+g^{2}NV_{12}D_{2}
\eeq
where
$ S_{20} $ is the 2 gluon  "free"  Schroedinger operator for the energy
$ 1-j $
\beq
S_{20}=j-1-\omega(1)-\omega(2)
\eeq
$ \omega(k) $ is the gluon Regge trajectory (see e.g [5]) and
$ V_{12} $ is the BFKL interaction with the kernel
\beq
V(1,2;1',2')=\frac{k_{1}^{2}{k'_{2}}^{2}+k_{2}^{2}{k'_{1}}^{2}}
{{k'_{1}}^{2}{k'_{2}}^{2}(k_{1}-k'_{1})^{2}}-\frac{(k_{1}+k_{2})^{2}}
{{k'_{1}}^{2}{k'_{2}}^{2}}
\eeq

Our aim in this section is to demonstrate that, to the leading order in
$ N $, the 4 gluon amplitude is reduced to the 2 gluon amplitude
$ D_{2} $, which is asumed to be known as the solution to (4).
However before we go to  4 gluons  we have to study the 3 gluon case.

\subsection{The 3 gluon system}
Still at this level no simplifications occur in the high-colour limit,
so that the results can be borrowed from [1,2]. Three gluons may
combine either into a fully symmetric colourless state (
$ d$ -coupling) or into a fully antisymmetric one (
$ f $-coupling). In the high colour limit it is more convenient to use
their combination
\beq
h_{a_{1}a_{2}a_{3}}=d_{a_{1}a_{2}a_{3}}+if_{a_{1}a_{2}a_{3}}
\eeq
cyclic symmetric in the gluons 1,2 and 3 and
with the properties
\[
h^{*}_{a_{1}a_{2}a_{3}}=h_{a_{2}a_{1}a_{3}} \]\[
\sum_{cd}h^{*}_{acd}h_{bcd} =\delta_{ab} 2N(1-2/N^{2})\]\beq
\sum_{cd}h_{acd}h_{bcd} =-\delta_{ab}(4/N)
\eeq
Normalizing, we introduce two colour wave functions
\beq
|123\rangle=\frac{1}{\sqrt{2N^{3}}}h_{a_{1}a_{2}a_{3}},\  \
  |213\rangle=\frac{1}{\sqrt{2N^{3}}}h_{a_{2}a_{1}a_{3}}
\eeq
which are orthogonal in the limit
$ N\rightarrow\infty $, due to (8). They correspond to the three gluons
lying on the surface of a cylinder in a sequence 123 or 213 in the
direction of the quark loop. Since each pair of the
gluons is in the adjoint representation, the interaction
will not change the colour configuration. So  the equations for the 3
gluon amplitude in the  two colour states $ |123\rangle $ and
$ |213\rangle $ decouple.

To set up these equations we need their zero-order terms. They are
provided by the quark loop with 3 gluons attached to it in all possible
ways (8 diagrams). The colour factor is given by the trace
\[ \Tr\{t_{a_{1}}t_{a_{2}}t_{a_{3}}\}=(1/8)h_{a_{1}a_{2}a_{3}}\]
if the gluons run in the order 123 along the quark line. The rest
 of the contribution is given by the same function $f
$  as in (1), in which the momenta of the gluons attached to the
same line (i.e the quark line or the antiquark line) sum together
and the overall sign is given by
$ (-1)^{n_{q}} $ where
$ n_{q} $ is the number of gluons attached to the quark line.
Projecting onto the states (9) we find
\beq
D_{30}^{(123)}=-D_{30}^{(213)}=g\sqrt{N/8}(D_{20}(2)-D_{20}(1)-D_{20}(3))
\eeq

In the 3 gluon equation, apart from the inhomogeneous term (10), an 
aditional contribution appears coming from possible transitions from 2 
to 3 gluons (Fig. 1). These transitions are accomplished by the vertex 
$ K_{2\rightarrow 3} $ introduced by J.Bartels [6]:
\beq
K_{2\rightarrow 3}=ig^{3}f^{a'_{1}a_{1}c}f^{ca_{2}d}f^{da_{3}a'_{3}}
W(1,2,3;1'3')
\eeq
for the transition of two gluons ($k'_{1},a'_{1}$) and  (
$ k'_{3},a'_{3} $) into three gluons (
$ k_{1},a_{1} $), (
$ k_{2},a_{2} $) and (
$ k_{3},a_{3} $). The kernel 
$ W $ is symmetric in the gluons 1 and 3 and can be expresed through  
the BFKL kernel  as \beq 
W(1,2,3;1'3')=V(2,3;1'-1,3')-V(1+2,3;1'3') \eeq
In the following one of the pairs of final gluons 12 or 23 will have a 
definite total colour. As a result the prefactor in (11) can be 
simplified to
\[ig^{3}T^{(12)}\delta_{a_{1}a'_{1}} f^{a_{2}a_{3}a'_{3}}\]
or to
\[ig^{3}T^{(23)}\delta_{a_{3}a'_{3}} f^{a'_{1}a_{1}a'_{2}}\]
for the two mentioned cases, respectively, where 
$ T $ is the total colour in each case (
$-N $ for the colourless state and 
$ -N/2 $ for the adjoint state). 
Projecting onto the state 
$ |123\rangle $  we find a contribution
\beq
D_{2\rightarrow 3}^{(123)}=g^{3}\sqrt{N^{3}/8}W(1,2,3;1'3')\otimes
D_{2}(1',3')\equiv g^{3}\sqrt{N^{3}/8}W_{2}(1,2,3)
\eeq
Here the symbol 
$ \otimes$ means integration over primed variables with the weight
$ (2\pi)^{-3}\delta^{2}(1+2+3-1'-3') $. This integration and the 
function 
$ D_{2}(1',3') $ are implicit in the abbreviated notation in the second 
equality (the subindex 2 shows that it is the function 
$ D_{2} $ which should be integrated with the kernel 
$ W $).

The complete equation for the 
$ |123\rangle $ colour state is thus
\beq
S_{30}D_{3}^{(123)}=D_{30}^{(123)}+D_{2\rightarrow 3}^{(123)}+
(1/2)g^{2}N(V_{12}+V_{23}+V_{31})D_{3}
\eeq
The operator 
$ S_{30} $ is a natural generalization of (5) to 3 gluons. The factor 
1/2 in front of the interaction part is due to the fact that all three 
gluons are now in the adjoint representation together with their 
neighbours.

The equation for  the 
$ |213\rangle $ state has opposite signs for its inhomogeneous parts. 
Therefore  we find
\beq
D_{3}^{(213)}=-D_{3}^{(123)}
\eeq

The seemingly complicated equation (14) can however be easily solved. 
As shown in [1] its full solution is simply the  zero-order term with 
all $ D_{20} $ substituted by the BFKL pomerons $ D_{2}$ ("the 
reggeized zero order term"):               \beq 
D_{3}^{(123)}=-D_{3}^{(213)}=g\sqrt{N/8}(D_{2}(2)-D_{2}(1)-D_{2}(3))
\eeq
This fact can be demonstrated by a direct substitution. We shall 
derive it in a different manner, using the idea that a pair 
of reggeized gluons in the adjoint representation  combines in a single 
reggeized gluon, presented in [4]. This allows to easily generalize
the derivation to the 4 gluon case.

Consider a 3 gluon equation with a zero-order term 
$ D_{2}{(1)} $
\beq
S_{30}D_{3}^{(1)}=D_{20}(1)+F_{3}^{(1)}+(1/2)g^{2}N(V_{12}+V_{23}+V_{31})
D_{3}^{(1)}
\eeq
with a certain additional inhomogeneous term $F_{3}^{(1)}$.
As shown in [4], the solution to (17) will be given by
\beq D_{3}^{(1)}=D_{2}(1) \eeq
provided we choose
\beq
F_{3}^{(1)}=-(1/2)g^{2}N (W_{2}(2,3,1)
      + W_{2}(3,2,1))
\eeq
(see definition (13)).

Indeed, put (18) into Eq. (17). The interaction 
$ V_{23} $ gives
\[(1/2)g^{2}NV_{23}D_{2}(1)=(\omega(2+3)-\omega(2)-\omega(3))D_{2}(1) \]
( "the bootstrap relation", [7]) and thus converts the operator
$ S_{30}(1,2,3) $ into $S_{20}(1,2+3)$. The two other interaction terms
give explicitly
\beq (1/2)g^{2}N(V(1,2;1',2')\otimes D_{2}(1',2'+3)+
V(3,1;3',1')\otimes D_{2}(1',2+3'))\eeq
On the other hand, using (12), the term 
$ F_{3}^{(1)} $ has the form
\[
F_{3}^{(1)}=-(1/2)g^{2}N(V(3,1;2'-2,1')\otimes D_{2}(2',1')-
V(2+3,1;2',1')\otimes D_{2}(2',1')
\]\beq+V(2,1;3'-3,1')\otimes D_{2}(3',1')-
V(3+2,1;3',1')\otimes D_{2}(3',1'))
\eeq
Taking as integration variables 
$ 3'=2'-2 $ in the first term in (21) and 
$ 2'=3'-3 $ in the third, we find that these terms cancel the second 
and first terms in (20), respectively. The second and fourth terms in 
(21) give in the sum \[ g^{2}NV(1,2+3;1',2')\otimes D_{2}(1',2')\] so 
that the equation (17) becomes \[ 
S_{20}(1,2+3)D_{2}(1,2+3)=D_{20}(1,2+3)+g^{2}NV_{1,2+3}D_{2}(1,2+3)\] 
which is evidently true.

The described procedure corresponds to combining  two reggeized 
gluons 2 and 3 in the adjoint representation into a single reggeized 
gluon with the momentum 2+3. For it to work, it is evidently necessary
that the zero order term be a function of only the sum 2+3 and also 
that a certain inhomogeneous term (19) be present which
describes the necessary contributions from the transitions from 2 to 
3 gluons.

Now we repeat this exercise with zero-order terms 
$ D_{20}(2) $ and 
$ D_{20}(3) $  and find two more solutions 
$ D_{3}^{(2)}=D_{2}(2) $ and 
$ D_{3}^{(3)}=D_{2}(3) $ which require additional inhomogeneous terms
$ F_{3}^{(2)} $ and 
$ F_{3}^{(3)} $, respectively, obtained from (19) by permutations of 123
to 231 and 312. Taking the combination (16) of these solutions we find 
that the additional inhomogeneous term to be added is
\[ g\sqrt{N/8}(F_{3}^{(2)}-F_{3}^{(1)}-F_{3}^{(3)}) =g^{3}\sqrt{N^{3}/8}
W_{2}(1,2,3)\]
which is precisely the term we have in  Eq. (14) for the 3-gluon 
amplitude. Therefore (16) is indeed its solution.
 \subsection{The 4-gluon system}

On the 4 gluon level simplifications due to 
$ N\rightarrow\infty $ become essential. At zero-order the 4 gluon 
amplitude is given by the quark loop with 4 gluons attached to it in 
all possible ways (16 diagrams). For a given order of gluons along the 
quark line, say, 1234,  the cyclic symmetric colour trace is
\beq
\Tr\{t_{a_{1}}t_{a_{2}}t_{a_{3}}t_{a_{4}}\}=
(1/8)h_{a_{1}a_{2}b}h_{a_{3}a_{4}b}+(1/4N)\delta_{a_{1}a_{2}}
\delta_{a_{3}a_{4}}
\eeq
At 
$ N\rightarrow\infty $ we neglect the second term and the sum
$h_{a_{1}a_{2}b}h_{a_{3}a_{4}b}$ becomes cyclic symmetric in the 
gluons. It defines a  colour state of 4 gluons
\beq
|1234\rangle =\frac{1}{2N^{2}}h_{a_{1}a_{2}b}h_{a_{3}a_{4}b}
\eeq
corresponding to the gluons lying on the surface of a cylinder in the 
order 1234 and in the adjoint representation with their neighbours. 
Other states are obtained from this by permutations of 234. In the 
limit 
$ N\rightarrow\infty $ they are orthonormalized. Projecting the quark 
loop onto these states we find
\beq
D_{40}^{(1234)}=D_{20}^{(4321)}=(1/4)g^{2}N(D_{20}(1)+D_{20}(4)-D_{20}(1+4))
\eeq 
\beq
D_{40}^{(2134)}=D_{20}^{(4312)}=(1/4)g^{2}N(D_{20}(2)+D_{20}(3)-D_{20}(1+2)-
D_{20}(1+3))
\eeq 

Interactions between the gluons do not change their overall colour 
state.  Indeed, interactions between neighbour gluons are diagonal in 
colour since neighbours are all in the adjoint representation. 
Interactions across the cylinder (e.g. 13 in the state $ |1234\rangle 
$) are down by $ 1/N $ and can be neglected in the leading 
approximation. Therefore at $ N\rightarrow\infty $ the complicated 
Bartels system for the 4 gluon amplitudes reduces to 4 decoupled 
equations for colour configurations appearing in (24) and (25). 
Moreover, one finds that the amplitudes with the inverse order of 
gluons (e.g. 1234 and 4321) are equal. As a result we arrive at only 
two different equations for colour states $ |1234\rangle $ and $ 
|2134\rangle $.

To set up these equations we need to find contributions from 
transitions into the 4 gluon state from 2- and 3-gluon states. 
The transition from 2 to 4 gluons corresponds to a single diagram, 
shown in Fig. 2. It is accomplished by a vertex similar to (11)
\beq
K_{2\rightarrow 4}=-g^{2}f^{a'_{1}a_{1}c}f^{ca_{2}d}f^{da_{3}e}
f^{ea_{4}a'_{4}}W(1,2+3,4;1',4')
\eeq
If gluons 12 and 34 have a definite total colour 
$ T $ the prefactor in (26) simpljfies to
\[ -g^{4}T^{(12)}T^{(34)}\delta_{a_{1}a'_{1}}\delta_{a_{2}a_{3}}
\delta_{a_{4}a'_{4}}\]
For the amplitudes in the states indicated in (24) and (25) we have 
$ T^{(12)}=T^{(34)}=-N/2 $. Projecting (26) onto these states we obtain
the contribution to the state $|1234\rangle$ (and 
$ |4321\rangle $)
\beq
D_{2\rightarrow 4}^{(1234)}=-(1/4)g^{4}N^{2}W_{2}(1,2+3,4)
\eeq
The contribution from the transition 2 to 4 to the states 
$ |2134\rangle $ and 
$ |4312\rangle $ turns out to be zero.

Transitions from 3 to 4 gluons are described by four diagrams shown in 
Fig. 3 and are accomplished by the vertex 
$ K_{2\rightarrow 3} $  described above. Projecting onto the states 
$ |1234\rangle $ and 
$ |2134\rangle $ we find the following contributions. For 
$ |1234\rangle$ 
\beq
D_{3\rightarrow 4}^{(1234)}=g^{3}\sqrt{N^{3}/8}(W(2,3,4;2',4')\otimes
D_{3}^{(124)}(1,2',4')+W(1,2,3;1',3')\otimes D_{3}^{(134)}(1',3',4))
\eeq
and for 
$ |2134\rangle $
\beq
D_{3\rightarrow 4}^{(2134)}=-g^{3}\sqrt{N^{3}/8}(W(1,2,4;1',4')\otimes
D_{3}^{(134)}(1',3,4')+W(1,3,4;1',4')\otimes D_{3}^{(124)}(1',2,4'))
\eeq

Thus we find the  equations for the two independent 4 gluon amplitudes
in the form
\beq
S_{40}D_{4}^{(1234)}=D_{40}^{(1234)}+D_{2\rightarrow 4}^{(1234)}+
D_{3\rightarrow 4}^{(1234)}+(1/2)g^{2}N(V_{12}+V_{23}+V_{34}+V_{41})
D_{4}^{(1234)}
\eeq
and
\beq
S_{40}D_{4}^{(2134)}=D_{40}^{(2134)}+
D_{3\rightarrow 4}^{(2134)}+(1/2)g^{2}N(V_{21}+V_{13}+V_{34}+V_{42})
D_{4}^{(2134)}
\eeq
We are going to demonstrate that solutions to these equations are 
exactly given by their zero-order terms 
$ D_{40} $ in which 
$ D_{20} $ are substituted by the BFKL pomerons 
$ D_{2} $, that is
\beq
D_{4}^{(1234)}=(1/4)g^{2}N(D_{2}(1)+D_{2}(4)-D_{2}(1+4))
\eeq 
and
\beq
D_{4}^{(2134)}=(1/4)g^{2}N(D_{2}(2)+D_{2}(3)-D_{2}(1+2)-
D_{2}(1+3))
\eeq 

To do it we apply the same trick as used in the 3 gluon case. Namely we 
shall find inhomogeneous terms necessary to combine 4 gluons into 2 for 
each term in the zero order expressions (24) and (25) separately. Then 
we shall sum the thus constructed pomerons into the combinations (32) 
and (33) and check that the resulting inhomogeneous terms will coincide 
with the terms $ D_{2\rightarrow 4} $  and $ D_{3\rightarrow 4} $ in 
Eqs. (30) and (31). Since the derivation closely follows the 3 gluon 
case, we omit all the details and discuss only the inhomogeneous terms 
needed at each step.

Let us start with the colour state 
$ |1234\rangle $. First consider an equation with a zero-order term
$ D_{20}(1)\equiv D_{20}(1,2+3+4) $. If we want to solve it by 
$ D_{4}^{(1)}=D_{2}(1,2+3+4) $ the three gluons 2,3 and 4 have to 
combine into a single one with the momentum 
$ k_{2}+k_{3}+k_{4} $. To achieve that we first combine the gluons 2 
and 3. According to our prescription explained above, this needs an 
inhomogeneous term
\beq
-(1/2)g^{2}N (W(2,3,4;2',4')\otimes D_{3}^{(1)}(1,2',4')+
W(3,2,1;3',1')\otimes D_{3}^{(1)}(1',3',4))
\eeq
where 
$ D_{3}^{(1)}(1,2,3)=D_{2}(1,2+3) $. Then we want the gluon 2+3 to 
combine with gluon 4. This will add a new inhomogeneous term
\beq
-(1/2)g^{2}N(W_{2}(2+3,4,1)+W_{2}(4,2+3,1))
\eeq
With these inhomogeneous terms the solution to the 4 gluon equation
can be easily shown to be 
$ D_{4}^{(1)}=D_{3}^{(1)}(1,2+3,4)=D_{2}(1,2+3+4) $. To abbreviate we 
 denote
\beq
W(2,3,4;2',4')\otimes D_{3}^{(1)}(1,2',4')\equiv W_{3}^{(1)}(2,3,4)=
W_{3}^{(1)}(4,3,2)
\eeq
\beq
W(3,2,1;3',1')\otimes D_{3}^{(1)}(1',3',4)\equiv W_{3}^{(1)}(3,2,1)=
W_{3}^{(1)}(1,2,3)
\eeq
(the superscript 
$ (1) $ denotes the distinguished gluon in the three gluon amplitude:
$ D_{3}^{(1)}(1,2',4)=D_{2}(1,2'+4) $ etc.). With this notation the 
 total inhomogeneous term added for the solution 
$ D_{4}^{(1)} $ is
\beq F_{4}^{(1)}=-(1/2)g^{2}N (W_{3}^{(1)}(2,3,4)+W_{3}^{(1)}(3,2,1)+
W_{2}(2+3,4,1)+W_{2}(4,2+3,1))\eeq
Likewise for the solution 
$ D_{4}^{(4)}=D_{2}(4,1+2+3) $ the necessary inhomogeneous term is
\beq F_{4}^{(4)}=-(1/2)g^{2}N (W_{3}^{(4)}(2,3,4)+W_{3}^{(4)}(3,2,1)+
W_{2}(2+3,1,4)+W_{2}(1,2+3,4))\eeq

Finally we study the solution 
$ D_{4}^{(14)}=D_{2}(1+4,2+3) $ . To construct it we first combine the 
gluons 23 and afterwards the gluons 14. The necessary inhomogeneous 
term
is found to be
\beq F_{14}^{(1)}=-(1/2)g^{2}N (W_{3}^{(2)}(2,3,4)+W_{3}^{(3)}(3,2,1)+
W_{2}(4,1,2+3)+W_{2}(1,4,2+3))\eeq

Now we combine the solutions 
$ D_{4}^{(1)} $, 
$ D_{4}^{(4)} $ and 
$ D_{4}^{(14)} $ into the combination (32). The resulting inhomogeneous 
term is
\[
F_{4}=-(1/8)g^{4}N^{2}(W_{3}^{(1)}(2,3,4))
+W_{3}^{(4)}(2,3,4))-W_{3}^{(2)}(2,3,4))\]\beq +
W_{3}^{(1)}(3,2,1)+W_{4}^{(4)}(3,2,1)-W_{3}^{(3)}(3,2,1)
+2W_{2}(1,2+3,4))
\eeq
Compare this with the inhomogeneous terms 
$ D_{2\rightarrow 4} $ and 
$ D_{3\rightarrow 4} $ in Eq. (30). The last term in (41) evidently 
coincides with 
$ D_{2\rightarrow 4} $. Putting the solution 
$ D_{3}^{(123)} $ into the term 
$ D_{3\rightarrow 4} $ we find a sum of functions 
$ W_{3} $ which exactly coincides with an analogous sum in (41). Thus 
we find that (32) is indeed the solution to Eq. (30).  

Now we pass to the colour state 
$ |2134\rangle $. Again we first construct four solutions 
$ D_{4} ^{(2)}=D_{2}(2,1+3+4) $, $D_{4}^{(3)}=D_{2}(3,1+2+4)$,
$ D_{4}^{(12)}=D_{2}(1+2,3+4) $ and 
$ D_{4}^{(13)}=D_{2}(1+3,2+4) $. Additional inhomogeneous terms 
necessary to combine 4 gluons into 2 gluons are found to be
(symmetrized in the two possible orders of this procedure)
\[
F_{4}^{(2)}=-(1/4)g^{2}N(W_{3}^{(2)}(3,4,2)+2W_{3}^{(2)}(1,3,4)+
W_{3}^{(2)}(3,1,2)+\]\beq W_{2}(1,3+4,2)+W_{2}(3+4,1,2)+W_{2}(4,1+3,2)+
W_{2}(1+3,4,2))
\eeq
\[
F_{4}^{(3)}=-(1/4)g^{2}N(W_{3}^{(3)}(2,1,3)+2W_{3}^{(3)}(1,2,4)+
W_{3}^{(3)}(2,4,3)+\]\beq W_{2}(1,2+4,3)+W_{2}(2+4,1,3)+W_{2}(4,1+2,3)+
W_{2}(1+2,4,3))
\eeq
\[
F_{4}^{(12)}=-(1/4)g^{2}N(W_{3}^{(3)}(3,4,2)+W_{3}^{(4)}(1,3,4)+
W_{3}^{(1)}(1,2,4)+\]\beq
W_{3}^{(2)}(3,1,2)+W_{2}(3,4,1+2)+W_{2}(3+4,1,2)+W_{2}(4,3,1+2)+
W_{2}(1,2,3+4))
\eeq
\[
F_{4}^{(13)}=-(1/4)g^{2}N(W_{3}^{(2)}(3,4,2)+W_{3}^{(1)}(1,3,4)+
W_{3}^{(4)}(1,2,4)+\]\beq
W_{3}^{(3)}(3,1,2)+W_{2}(2,4,1+3)+W_{2}(2+4,1,3)+W_{2}(1,3,2+4)+
W_{2}(4,2,1+3))
\eeq
Forming the combination (33) we find the total inhomogeneous term
\[
F_{4}=(1/2)g^{2}N(F_{4}^{(2)}+F_{4}^{(3)}-F_{4}^{(12)}-F_{4}^{(13)})=\]\[
-(1/16)g^{4}N^{2}(2W_{3}^{(2)}(1,3,4)+2W_{3}^{(3)}(1,2,4)-
W_{3}^{(1)}(1,2,4)-W_{3}^{(1)}(1,3,4)-\]\[ W_{3}^{(4)}(1,2,4)-W_{3}^{(4)}
(1,3,4)+W_{2}(2,1+3,4)+W_{2}(1,3+4,2)+W_{2}(3,1+2,4)+\]
\beq W_{2}(1,2+4,3)-
W_{2}(1+2,3,4)-W_{2}(1,2,3+4)-W_{2}(4,2,1+3)-W_{2}(1,3,2+4))
\eeq
On the other hand, the existing inhomogeneous term (29) can be rewritten
using (16) as
\[
D_{3\rightarrow 4}^{(2134)}=-(1/8)g^{4}N^{2}(W_{3}^{(2)}(1,3,4)-
W_{3}^{(1)}(1,3,4)-W_{3}^{(4)}(1,3,4)+ \]\beq
W_{3}^{(3)}(1,2,4)-W_{3}^{(1)}(1,2,4)-W_{3}^{(4)}(1,2,4))
\eeq
At first sight they are different, since terms 
$ 3\rightarrow 4$ do not coincide and terms 
$ 2\rightarrow 4 $ are absent in (47). However,
functions  of the type
$ W_{3}^{(1)}(1,3,4) $ can in fact  be expressed in terms of 
$ W_{2} $.

 Indeed we have, by definition   and using (12),
\beq
W_{3}^{(1)}(1,3,4)=(V(3,1;4'-4,1')-V(3+4,1;4',1'))\otimes D_{2}(1',2+4')
\eeq
Shifting the integration variables we have
\[ V(3,1;4'-4,1')\otimes D_{2}(1',2+4')=V(3,1;3'-2-4,1')\otimes 
D_{2}(1',3')=\]\[
 W_{2}(2+4,3,1)+V(2+3+4,1;3',1')\otimes D_{2}(3',1')\]
\[V(3+4,1;4',1')\otimes D_{2}(1',2+4')=V(3+4,1;2'-2,1')\otimes 
D_{2}(1',2')=\]\[ W_{2}(2,3+4,1)+V(2+3+4,1;3',1')\otimes D_{2}(3',1')\]
Subtracting we get an identity
\beq
W_{3}^{(1)}(1,3,4)=W_{2}(2+4,3,1)-W_{2}(2,3+4,1)
\eeq
and interchanging 1 and 4
\beq
W_{3}^{(4)}(1,3,4)=W_{2}(1+2,3,4)-W_{2}(2,1+3,4)
\eeq

Using these identities we find that all terms with 
$ W_{2} $  in (46) combine into
\[-W_{3}^{(1)}(1,2,4)-W_{3}^{(1)}(1,3,4)-W_{3}^{(4)}(1,2,4)-W_{3}^{(4)}
(1,3,4)\]
which together with the rest of the terms makes (46) identical to (47).

Thus we have proven that in the leading order in 
$ 1/N $ the solution to the 4 gluon equation is exactly given by 
(32) and (33) and thus reduces to single BFKL pomerons.

\section{Next-to-leading order in 
$ 1/N $}
\subsection{Equation for the two-pomeron amplitude}

At next-to-leading order the colour classification of 4 gluon states 
looses its cyclic symmetry, so that one has to return to the standard 
description in terms of colour states of the two subsystems of gluons 
12 and 34. We shall be interested in a state which is a color 
singlet in both subsystems, with a colour wave function
\beq
|0\rangle=(1/N^{2})\delta_{a_{1}a_{2}}\delta_{a_{3}a_{4}}
\eeq
This state is absent in the leading approximation in 
$ 1/N $ and first appears at the next-to-leading order. It is this 
state that corresponds to the coupling of the projectile with two
pomerons and enters the diffractive processes. Projecting the quark 
loop onto the state (51) we find the corresponding zero-order amplitude:
\beq
D_{40}^{(0)}=(1/2)g^{2}(D_{20}(1)+D_{20}(2)+D_{20}(3)+D_{20}(4)-
D_{20}(1+2)-D_{20}(1+3)-D_{20}(1+4))
\eeq
It is evidently down by a factor 
$ 1/N $ as compared to the leading order amplitudes 
$ D_{4}^{(1234)} $ and 
$ D_{4}^{(2134)} $.

The interaction between gluons in the state 
$ |0\rangle$ may lead them either to the same state
(terms with 
$ V_{12} $ and 
$ V_{34} $) or to the state in which subsystems 12  and 34 are both 
in the antisymmetric adjoint representation (the 
$ f$  part of $ h $). The amplitude for the latter state is already 
known (in the leading order in 
$ 1/N $). Therefore  transitions from the 4 gluons in the adjoint state
into state 
$ |0\rangle $ will serve as an additional inhomogeneous term 
$ D_{4\rightarrow 4}^{(0)} $ in the equation for the amplitude 
$ D_{4}^{(0)} $. Since no other colour state will enter in the 
next-to-leading order, the resulting equation for 
$ D_{4}^{(0)} $ will again decouple and have the form
\beq
S_{40}D_{4}^{(0)}=D_{40}+D_{2\rightarrow 4}^{(0)}+D_{3\rightarrow 
4}^{(0)}+D_{4\rightarrow 4}^{(0)}+g^{2}N(V_{12}+V_{34})D_{4}^{(0)}
\eeq

The zero-order term 
$ D_{4}^{(0)} $ is given by (52). The contributions from the 
transitions $2\rightarrow 4 $ and $ 3\rightarrow 4 $ are described by 
the same diagrams of Fig. 2 and 3, although with new colour factors. 
One easily finds them to be \beq D_{2\rightarrow 
4}^{(0)}=-g^{4}NW_{2}(1,2+3,4) \eeq and \[ D_{3\rightarrow 
4}^{(0)}=g^{3}\sqrt{2N}(W(1,2,3;1'3')\otimes 
D_{3}^{(134)}(1',3',4)-W(1,2,4;1',4')\otimes D_{3}^{(134)}(1',3,4')+\]\beq
W(2,3,4;2',4')\otimes D_{3}^{(124)}(1,2',4)-W(1,3,4;1',4')\otimes
D_{3}^{(124)}(1',2,4'))
\eeq
where the amplitude 
$ D_{3}^{(123)} $  is given by (16).

The terms 
$ 4\rightarrow 4 $ come from 4 leading colour states 
$ |1234\rangle $, 
$ |2134\rangle $,
$ |4321\rangle $ and
$ |4312\rangle $ as a result of interactions of gluons belonging to 
different subsystems, that is upon action of 
$ K_{2\rightarrow 2}^{(13)} $,
$ K_{2\rightarrow 2}^{(14)} $,
$ K_{2\rightarrow 2}^{(23)} $ and
$ K_{2\rightarrow 2}^{(13)} $, where, for example
\beq
 K_{2\rightarrow 2}^{(13)}=-g^{2}(T_{1}T_{3})V_{13}
\eeq
Contributions from interactions of gluons across the cylinder (e.g of 
1 and 3 in the configuration 
$ |1234\rangle $) are calculated trivially and we obtain in this manner 
a part of 
$ D_{4\rightarrow 4}^{(0)} $
\beq
-g^{2}((V_{13}+V_{24})D_{4}^{(1234)}+(V_{23}+V_{14})D_{4}^{(2134)})\eeq

Calculating the contribution from the interaction between adjacent 
gluons needs some care, since its leading order result is evidently 
zero. Suppose we want to find 
$ \langle 1234|K_{2\rightarrow 2}^{(23)}|0\rangle $.
For the initial state we can take just as well a cyclic symmetric 
combination
\[(1/2N^{2})(h_{a_{1}a_{2}b}h_{a_{3}a_{4}b}+(2/N)\delta_{a_{1}a_{2}}
\delta_{a_{3}a_{4}})\]
since all (12)(34) states except the antisymmetric adjoint one give 
zero.
Using the cyclic symmetry we rewrite it as
\[(1/2N^{2})(h_{a_{2}a_{3}b}h_{a_{4}a_{1}b}+(2/N)\delta_{a_{2}a_{3}}
\delta_{a_{4}a_{1}})\]
The first term corresponds to gluons 23 in the adjoint 
representation, the second - in the colourless one. Acting on this, 
operator 
$ -(T_{2}T_{3}) $ gives
\[(1/2N^{2})((N/2)h_{a_{2}a_{3}b}h_{a_{4}a_{1}b}+N(2/N)\delta_{a_{2}a_{3}}
\delta_{a_{4}a_{1}})=\]\[
(1/4N)h_{a_{1}a_{2}b}h_{a_{3}a_{4}b}+(1/2N^{2})(\delta_{a_{1}a_{2}}
\delta_{a_{3}a_{4}}+\delta_{a_{2}a_{3}}
\delta_{a_{4}a_{1}})\]
Projecting  onto the state 
$ |0\rangle $ we finally obtain
\[ \langle 1234|K_{2\rightarrow 2}^{(23)}|0\rangle =
(1/2)g^{2}V_{23}\]
An identical contribution comes from the state 
$ |4321\rangle $.

 Calculating in this manner all adjacent interactions 
we find the second half of 
$ D_{4\rightarrow 4}^{(0)} $;
\beq
g^{2}((V_{23}+V_{14})D_{4}^{(1234)} +(V_{13}+V_{24})D_{4}^{(2134)})
\eeq
Summing (57) and (58) we finaly find
\beq
D_{4\rightarrow 
4}^{(0)}=g^{2}(V_{23}+V_{14}-V_{13}-V_{24})(D_{4}^{(1234)}-
D_{4}^{(2134)})
\eeq

In the equation (53) the three inhomogeneous terms 
$ D_{2\rightarrow 4}^{(0)} $, 
$ D_{3\rightarrow 4}^{(0)} $ and 
$ D_{4\rightarrow 4}^{(0)} $  are each the result of some operator 
applied to the BFKL pomeron, since  amplitudes 
$ D_{3} $ and 
$ D_{4} $ entering them are, in fact, linear combinations of BFKL 
pomerons. Therefore we can present 
\beq
 D_{2\rightarrow 4}^{(0)}+ 
 D_{3\rightarrow 4}^{(0)}+ 
 D_{4\rightarrow 4}^{(0)}=Z(1,2,3,4;1'2') \otimes D_{2}(1',2')
\eeq
The kernel 
$ Z $  describes the coupling of the initial pomeron attached to the 
quark loop (the projectile) to the two final ones. It is our 
three-pomeron vertex. In the following we shall present a more explicit
form of 
$ Z $.

 In terms of 
$ Z $ we write the equation for $D_{4}^{(0)}$ as follows
\beq
S_{40}D_{4}^{(0)}=D_{20}^{(0)}+Z\,D_{2}+g^{2}N(V_{12}+V_{34}) 
D_{4}^{(0)}
\eeq
Its solution may evidently be split into two parts: a direct one and 
the triple pomeron part, which separately satisfy this equation with 
the inhomogeneous terms 
$ D_{40}^{(0)} $ and 
$ Z\,D_{2} $, respectively. Both parts can easily be found 
explicitly due to the fact that the operator acting in (61) is 
evidently a sum of two independent parts for subsystems 12 and 34.

\subsection{The direct contribution}

To solve the equation for the direct part we represent the quark loop 
diagrams in (1) as a Fourier transform:
\beq
g^{2}Nf(k)=\int d^{2}r \rho(r)e^{ikr}
\eeq
Function 
$ \rho(r) $ evidently describes the colour density created by the 
$ q\bar q $ pair with  a transverse dimension 
$ r $. Using this representation one finds that the term 
$ D_{20}^{(0)} $  (Eq.(52) ) can be represented as
\beq
D_{20}^{(0)}=(1/4)g^{2}\int d^{2}r\rho(r)\prod_{j=1}^{4}(e^{ik_{j}r}-1)
\eeq
It thus factorizes in the gluons at fixed 
$ r $.

Accordingly we can start with an equation for the direct part at fixed 
$ r $:
\beq
S_{40}D_{4}^{(r)}=\prod_{j=1}^{4}(e^{ik_{j}r}-1)+g^{2}N 
(V_{12}+V_{34})D_{4}^{(r)}
\eeq
Evidently in this equation subsystems 12 and 34 decouple. The 
solution to (64) is given by the convolution in the "energy" 
$ 1-j $ of two independent BFKL pomerons
\beq
D_{4}^{(r)}=\int dj_{12} 
dj_{34}\delta(j-j_{12}-j_{34})D_{2,j_{12}}^{(r)}(1,2) 
D_{2,j_{34}}^{(r)}(3,4) \eeq Here the pomeron $ D_{2,j}^{(r)}(1,2) $
satisfies the equation
\beq
S_{20}D_{2,j}^{(r)}=\prod_{j=1}^{2}(e^{ik_{j}r}-1) 
+g^{2}NV_{12}D_{2,j}^{(r)}
\eeq
and similarly for the second pomeron.

The final direct amplitude will result upon integration over 
$ r $:
\beq
D_{4}^{(dir)}=(1/4)g^{2}\int d^{2}r \rho(r)D_{4}^{(r)}
\eeq
This contribution exactly corresponds to two pomerons directly coupled 
to the quark loop. As we observe, this coupling factorizes at fixed 
interquark distance. The corresponding amplitude has thus an eikonal 
form at fixed 
$ r $. This feature was first established in [4] where it was also 
generalized to any number of pomerons directly coupled to the quark 
loop.

\subsection{The three-pomeron vertex}
The triple pomeron part can be written in terms of the Green function
of Eq. (61) as
\beq
D_{4}^{(triple)}=G_{4}(1,2,3,4;1',2',3',4')\otimes 
Z(1',2',3',4';1'',2'')\otimes D_{2}(1'',2'')
\eeq
The Green function 
$ G_{4} $ is evidently a convolution in the energy of two two-gluon
(BFKL) Green functions
\beq
G_{4,j}(1,2,3,4;1',2',3',4')=\int dj_{12}dj_{34}\delta (j-j_{12}-j_{34})
G_{2,j_{12}}(1,2,;1'2')G_{2,j_{34}}(3,4;3',4')
\eeq
If the target part consists of two quark loops, analogous to the 
projectile loop, then the final solution will be given by two pomerons 
$ D_{2} $ for subsystems 12 and 34 attached to the three-pomeron vertex 
$ Z $ acting on the pomeron $D_{2}
$  coming from the projectile. 

 The three-pomeron vertex $ Z $ is implicitly 
determined by its definition (60), Eqs. (54), (55) and (59) for the 
terms $ D_{2\rightarrow 4} $, $ D_{3\rightarrow 4} $ and $ 
D_{4\rightarrow 4} $ and expressions (16), (32) and (33) for the 
leading order solutions $ D_{3} $ and $ D_{4} $. Our aim here is to 
obtain an explicit expression for $ Z $ which will facilitate the study 
of its infrared behaviour and comparison with the symmetric vertex 
introduced in [2] for $ N=3 $.

We start with the term 
$ \dtf^{(0)}$. Putting (16) into (55) we find terms of two types. Terms 
of the type $ W_{3}^{(3)}(1,2,3) $ can all be reduced to functions $ 
W_{2} $. Taking into account that only terms symmetric under the 
interchange 1 and 2 and/or 3 and 4 contribute, we find for this part
\beq
\dtf^{(01)} =(1/2)g^{4}N (W_{2}(1+4,2,3)+W(1+3,2,4)-2W(3,1+2,4)+\\
		   (1234\rightarrow 4321)
\eeq
The second part of 
$ \dtf^{(0)} $ is formed by  terms of the type 
$ W_{3}^{(1)}(2,3,4) $. Using (12) and the bootstrap relation  we find
\beq
(1/2)g^{2}NW_{3}^{(1)}(2,3,4)=D_{2}(1)(\omega(2+3)+\omega(3+4)
-\omega(2+3+4)-\omega(3))
\eeq
As a result, the second part of 
$ \dtf^{(0)} $ turns out to be
\[
\dtf^{(02)} =-g^{2}D_{2}(1)(\omega(2+3)+\omega(3+4)
-\omega(2+3+4)-\omega(3))\]\beq
-(1234\rightarrow 2134)-(1234\rightarrow 3124)-(1234\rightarrow 4123)
\eeq

To further simplify these expressions and also compare them with [2] we 
express functions 
$ W_{2} $ for the total momentum zero via the function 
$ G(k_{1},k_{2})=G(k_{2},k_{1}) $ introduced in [2]. For 
$ 1+2+3=0 $
\beq
g^{2}NW_{2}(1,2,3)=G(1+2,2+3)-D_{2}(1+2)(\omega(2)-\omega(1+2))-
D_{2}(2+3)(\omega(2)-\omega(2+3))
\eeq

In terms of functions 
$ G $, summing (70) and (72) we find
\[
\dtf^{(0)}=(1/2)g^{2}(G(1,2+4)+G(2,1+4)+G(3,1+4)+G(4,1+3)-
2G(1,2)-2G(3,4)-\]\[
D_{2}(1+3)(\omega(2)+\omega(3)-2\omega(1+3))-
D_{2}(1+4)(\omega(2)+\omega(3)-2\omega(1+4))+\]\[
D_{2}(1)(\omega(1)+\omega(3)-2\omega(1+4))+
D_{2}(2)(\omega(2)+\omega(3)-2\omega(1+3))+\]\beq
D_{2}(3)(\omega(2)+\omega(3)-2\omega(1+3))+
D_{2}(1)(\omega(2)+\omega(4)-2\omega(1+4))
) 
\eeq
This expression is not symmetric in gluons  12  nor in gluons 34. 
However the function 
$ D_{4}^{(0)} $ is symmetric in both gluon pairs. So we should 
symmetrize (74) in gluon pairs 12 and 34. The resulting symmetric 
expression is
\[
\dtf^{(0)}=(1/4)g^{2}(G(1,2+3)+G(1,2+4)+G(2,1+3)+G(2,1+4)\]\[
+G(3,1+4)+G(3,2+4)+G(4,1+3)+G(4,2+3)-
4G(1,2)-4G(3,4)-\]\[
D_{2}(1+3)(\sum_{i=1}^{4}\omega(i)-4\omega(1+3))-
D_{2}(1+4)(\sum_{i=1}^{4}\omega(i)-4\omega(1+4))+\]\[
D_{2}(1)(2\omega(1)+\omega(3)+\omega(4)-2\omega(1+3)-2\omega(1+4))+
D_{2}(2)(2\omega(2)+\omega(3)+\omega(4)-2\omega(1+3)-2\omega(1+4))+\]\beq
D_{2}(3)(2\omega(3)+\omega(1)+\omega(2)-2\omega(1+3)-2\omega(1+4))+
D_{2}(1)(2\omega(4)+\omega(1)+\omega(2)-2\omega(1+3)-2\omega(1+4))
) 
\eeq

 Note that in this notation the term $ \ddf^{(0)} $ can be 
rewritten as \beq 
\ddf^{(0)}=-g^{2}(G(1,4)+D_{2}(1)(\omega(1)-\omega(2+3))+
D_{2}(4)(\omega(4)-\omega(2+3)))
\eeq
After symmetrization in 12 and 34 it becomes
 \[ 
\ddf^{(0)}=-(1/4)g^{2}(G(1,3)+G(1,4)+G(2,3)+G(2,4)+\]\[
D_{2}(1)(2\omega(1)-\omega(1+3)-\omega(1+4))+
D_{2}(2)(2\omega(2)-\omega(1+3)-\omega(1+4))+ \]\beq
D_{2}(3)(2\omega(3)-\omega(1+3)-\omega(1+4))+
D_{2}(4)(2\omega(4)-\omega(1+3)-\omega(1+4)))
\eeq

We finally come to the term 
$ \dff^{(0)} $. We put expressions (32) and (33) for
$ D_{4}^{(1234)} $ and 
$ D_{4}^{(2134)} $ into (59).  Operator 
$ V_{23}+V_{14}-V_{13} -V_{24} $ is antisymmetric under the interchange 
of 1 and 2 and/or 3 and 4. Since the two-pomeron state is symmetric 
under these substitutions, only antisymmetric parts of 
$ D_{4}^{(1234)} $ and 
$ D_{4}^{(2134)} $ give a nonzero contribution. Taking this into 
account we find
\beq  
\dff^{(0)}=(1/4)g^{4}N (V_{23}+V_{14}-V_{13} -V_{24} )(D_{2}(1+3)-
D_{2}(1+4))
\eeq
Here we again meet with two differnt types of terms: with the 
interaction between different gluons (e.g 
$ V_{13} D_{2}(1+4) $) or within the same gluon (e.g. 
$ V_{13}D_{2}(1+3) $).

 The first type leads to a contribution
\beq
\dff^{(01)}=(1/4)g^{4}N((V_{13}+V_{24})D_{2}(1+4)+(V_{23}+V_{14})D_{2}(1+3))
\eeq
To reduce it to the expression of the same type as before we use an 
identity, which is proven in the Appendix
\beq
V_{13}D_{2}(1+4)=W_{2}(2+3,1,4)+W_{2}(2,3,1+4)-W_{2}(2,1+3,4)
-W_{2}(2+3,0,1+4)
\eeq
Introducing then functions 
$ G $ we find for this part
\[
\dff^{(01)}=(1/4)g^{2}(G(1,2+3)+G(1,2+4)+G(2,1+3)+\]\[
G(2,1+4)+G(3,1+4)+
G(3,2+4)+G(4,1+3)+G(4,2+3)-\]\[ G(1,3)-G(1,4)-G(2,3)-G(2,4)-
2G(2+3,1+4)-2G(1+3,2+4)+\]\[
D_{2}(1)(\omega(2+3)+\omega(2+4)-\omega(3)-\omega(4))+
D_{2}(2)(\omega(1+3)+\omega(1+4)-\omega(3)-\omega(4))+\]\[
D_{2}(3)(\omega(2+4)+\omega(1+4)-\omega(1)-\omega(2))+ 
D_{2}(4)(\omega(1+3)+\omega(2+3)-\omega(1)-\omega(2))- \]\beq
(D_{2}(1+3)+D_{2}(1+4))\sum_{i=1}^{4}\omega(i))
\eeq
The second part of the contribution is
\[
\dff^{(02)}=-(1/4)g^{4}N((V_{13}+V_{24})D_{2}(1+3)+
\]\beq (V_{23}+V_{14})D_{2}(1+4))
\eeq
It is much simpler. Application of the bootstrap relation  
immediately gives
	 \beq
\dff^{(02)}=(1/2)g^{2}N(D_{2}(1+4)(\sum_{i=1}^{4}\omega(i)-
2\omega(1+4))+
D_{2}(1+3)(\sum_{i=1}^{4}\omega(i)-2\omega(1+3)))
\eeq

The final three-pomeron vertex is obtained after summing (75), (77), 
(81) and (83). It is remarkable that all terms involving functions $ 
D_{2}(i) $ or $ D_{2}(i+k) $ cancel and the resulting expression 
contains only functions $ G $:  \[  
Z\,D_{2}=(1/2)g^{2}(G(1,2+3)+G(1,2+4)+G(2,1+3)+\]\[ G(2,1+4)+G(3,1+4)+ 
G(3,2+4)+G(4,1+3)+G(4,2+3)-\]
\beq 
2G(1,2)-2G(3,4)-G(1,3)-G(1,4)-G(2,3)-G(2,4)-
G(2+3,1+4)-G(1+3,2+4))
\eeq
The vertex 
$ Z $ itself can be easily extracted from this expression separating 
the operator which acts on 
$ D_{2} $. 

One immediately observes that the found vertex is infrared stable. It 
is expressed via functions 
$ G $, which are explicitly infrared stable by construction (see [2]).
Moreover one easily finds from (84) that the vertex 
$ Z $ goes to zero if any of its external variables (1,2,3 or 4 in (84))
goes to zero. This means that it can be safely integrated with two
pomerons $D_{2}(1,2)$ and 
$ D_{2}(3,4) $ which describe its coupling to the two target colourless
particles. By construction it is symmetric in gluons 12 and 34 and 
under the interchange  12 
$ \leftrightarrow $ 34. Thus it satisfies all requirements 
relevant from the physical point of view.  However, as one observes 
from (84), it is not fully symmetric in all the gluons
as the transitional vertex 
constructed in [2]. Related to this, it contains somewhat less terms 
than in [2] (16 instead of 19) and with different coefficients.
We shall discuss the physical reason for this difference in the next 
section.  

 \section{Discussion}

 The study of the 4-gluon system at 
$ N\rightarrow\infty $ leads to a very simple and physically expected
 picture. In the leading approximation the system reduces to a single 
 BFKL pomeron, the 4 gluons in the colour configuration of a cylinder 
 coalescing into a pair of gluons. In the next-to-leading order 
 the diffractive amplitude appears, which clearly divides into a direct 
 contribution (double pomeron exchange), with the two pomerons directly 
 coupled to the quark loop, and a triple pomeron contribution, with a 
 single pomeron splitting into two pomerons by a three-pomeron vertex.

 The found three-pomeron vertex is different from the one originally 
 constructed by J.Bartels and M.Wuesthoff in [2]. The reason is not 
 related to the limit 
$ N\rightarrow\infty $, as such, but to a different treatment of the 4-
 gluon amplitude in the colour state 
$ |0\rangle $. In [2] this amplitude was supposed to also contain a 
contribution from 2-gluon states (BFKL pomerons) in the form of the 
reggeized zero-order term (52). As a result in Eq. (53) new terms
$ 4\rightarrow 4 $ appear which desribe transitions from the reggeizing 
to non-reggeizing pieces in the same colour state 
$ |0\rangle $. This naturally changes the transitional vertex and 
explains the difference between between our vertex and that of [2].

However, in our opinion, the assumption about the existence of a 
reggeizing piece in the 4 gluon system in the state 
$ |0\rangle $ lacks any foundation. Reggeization, say, of the term 
$ D_{20}(1+2) $ in (52) implies the existence of a colourless reggeon,
identical to the physical one. Reggeization of the term 
$ D_{20}(1+3) $ implies the existence of such a reggeon in the 
symmetric representation of the highest dimension.  These 
possibilities are not supported by the BFKL equation for different 
colour states. The 
$ N\rightarrow\infty $ approach definitely rejects them.
In it the zero-order term (52) does not reggeize into a single pomeron 
but rather into two pomerons, thus leading to a well-defined and 
intuitively expected double pomeron exchange contribution.
For these reasons we consider our three-pomeron vertex (84) better
founded and suited to the decription of the diffractive amplitude, as 
the original one in [2].

\section{Acknowledgments}

The author expresses his deep gratitude to Prof. C.Pajares, who
attracted his attention to the problem, and to 
Prof. J.Bartels for 
illuminating discussions, which served as a starting point for this 
study.

\section{Appendix. Transformation of the term 
$ V_{13}D_{2}(1+4) $}

The term of interest is explicitly written as
 \beq
V(1,3;1',3')\otimes D_{2}(1'+4,2+3')=
V(1,3;1'-4,3'-2)\otimes D_{2}(1',2)=
V(1,3;4'-4,-4'-2)\otimes D_{2}(4',-4')
\eeq
In the last equality we used the conservation law 
$ 1+2+3+4=0 $. The identity (12) can be rewritten, again, with the 
conservation law taken into account, as
\beq
W(4,1,3;4',1')=V(1,3;4'-4,1+3+4-4')-V(1+4,3;4',1+3+4-4')
\eeq
Since 
$ 1+3+4=-2 $, integrating (86)  with 
$ D_{2}(4',-4') $ and comparing with (85) we find
\beq
V_{13}D_{2}(1+4)=(W(4,1,3;4',1+3+4-4')+V(1+4,3;4',1+3+4-4'))\otimes 
D_{2}(4',-4')
\eeq

Consider the two terms on the right-hand side separately. Interchanging
3 and 4 in the first we find (3'=1+3+4-4'):
\[
W(3,1,4:3',4')\otimes D_{2}(4',3'-1-3-4)=
(V(1,4;3'-3,4')
\]\beq-V(1+3,4,3',4'))\otimes D_{2}(4',3'-1-3-4)=
(V(1,4; 3''+1+4,4')-V(1+3,4,3''+1+3+4,4'))\otimes D_{2}(4',3'')
\eeq
where 3''=3'-1-3-4. Both terms can be expressed via 
$ W $, using (12). We obtain
\[
W(-1-4,1,4:3',4')+V(-4,4;3'4')-W(-1-3-4,1+3,4; 3',4')
\]\beq-V(-4,4;3',4'))
\otimes D_{2}(4',3')=W_{2}(-1-4,1,4)-W_{2}(-1-3-4,1+3,4)
\eeq
The second term in (87) is transformed analogously
\[
V(3,1=4;3',4')\otimes D_{2}(4',3'-1-3-4)=
V(3,1+4;3''+1+3+4,4')\otimes D_{2}(4',3'')=\]\beq
W_{2}(-1-3-4,3,1+4)+V(-1-4,1+4;3',4')\otimes D_{2}(4',3')
\eeq
Next we take into account that -1-4=2+3, -1-3-4=2 and that
\beq
V(2+3,1+4;3',4')\otimes D_{2}(4',3')=-W_{2}(2+3,0,1+4)
\eeq
Taking the sum of (89) and (90 we then obtain (80).

\newpage
\section{References}
1. J.Bartels, Z.Phys.{\bf C60} (1993) 471.\\
2. J.Bartels and M.Wuesthoff, Z.Phys. {\bf C66} (1995) 157.\\
3. J.Bartels, L.N.Lipatov and M.Wuesthoff, Nucl. Phys. {\bf B 464} 
(1996) 298.\\
4. M.A.Braun, Z.Phys. {\bf C71} (1996) 601.\\
5. L.N.Lipatov in: "Perturbative QCD", ed. A.H.Mueller, Advanced Series 
 in High Energy Physics (World Scientific, Singapore, 1989).\\
6. J.Bartels, Nucl. Phys. {\bf B175} (1980)365.\\
7. L.N.Lipatov, Yad. Fiz. {bf 23} (1976) 642.

\section{Figure captions}
Fig. 1. The transition from 2 to 3 gluons.\\
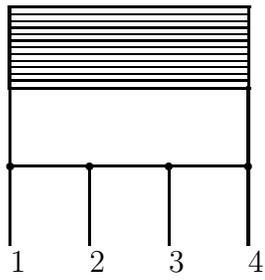
Fig. 2. The transition from 2 to 4 gluons.\\
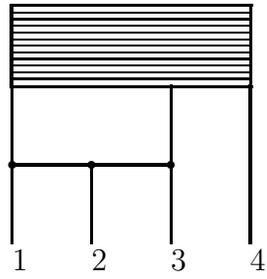
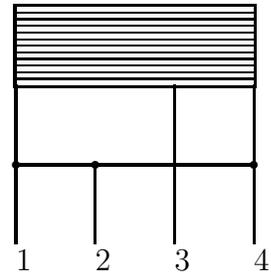
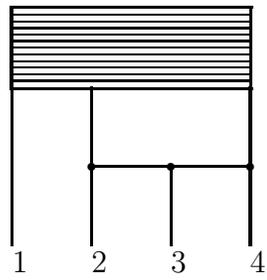
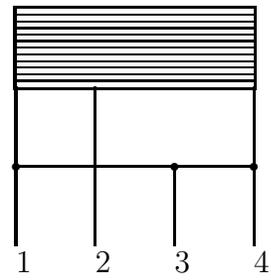
Fig. 3. The transition from 3 to 4 gluons.\\

\newpage
\section{Figures}
\begin{picture}(200,200)(-50,-50)
\thicklines
\put(50,50){\framebox (90,30)}
\put(50,50){\line (0,-1){60}}
\put(140,50){\line (0,-1){60}}
\put(50,20){\line (1,0){90}}
\put(95,20){\line (0,-1){30}}
\put(50,-20){\large 1}
\put(95,-20){\large 2}
\put(140,-20){\large 3}
\put(80,-50){\Large Fig. 1}
\put(50,20){\circle*{3}}
\put(95,20){\circle*{3}}
\put(140,20){\circle*{3}}
\thinlines
\multiput(50,52.5)(0,2.5){11} {\line(1,0){90}}
\end{picture}

\begin{picture}(200,200)(-50,-50)
\thicklines
\put(50,50){\framebox (90,30)}
\put(50,50){\line (0,-1){60}}
\put(140,50){\line (0,-1){60}}
\put(50,20){\line (1,0){90}}
\put(80,20){\line (0,-1){30}}
\put(110,20){\line (0,-1){30}}
\put(50,-20){\large 1}
\put(80,-20){\large 2}
\put(110,-20){\large 3}
\put(140,-20){\large 4}
\put(80,-50){\Large Fig. 2}
\put(50,20){\circle*{3}}
\put(80,20){\circle*{3}}
\put(110,20){\circle*{3}}
\put(140,20){\circle*{3}}
\thinlines
\multiput(50,52.5)(0,2.5){11} {\line(1,0){90}}
\end{picture}
\newpage
\begin{picture}(400,200)(0,0)
\put(0,0){\begin{picture}(200,200)(-50,-50)
\thicklines
\put(50,50){\framebox (90,30)}
\put(50,50){\line (0,-1){60}}
\put(140,50){\line (0,-1){60}}
\put(50,20){\line (1,0){60}}
\put(80,20){\line (0,-1){30}}
\put(110,50){\line (0,-1){60}}
\put(50,-20){\large 1}
\put(80,-20){\large 2}
\put(110,-20){\large 3}
\put(140,-20){\large 4}
\put(50,20){\circle*{3}}
\put(80,20){\circle*{3}}
\put(110,20){\circle*{3}}
\thinlines
\multiput(50,52.5)(0,2.5){11} {\line(1,0){90}}
\end{picture}}

\put(200,0){\begin{picture}(200,200)(-50,-50)
\thicklines
\put(50,50){\framebox (90,30)}
\put(50,50){\line (0,-1){60}}
\put(140,50){\line (0,-1){60}}
\put(50,20){\line (1,0){90}}
\put(80,20){\line (0,-1){30}}
\put(110,50){\line (0,-1){60}}
\put(50,-20){\large 1}
\put(80,-20){\large 2}
\put(110,-20){\large 3}
\put(140,-20){\large 4}
\put(50,20){\circle*{3}}
\put(80,20){\circle*{3}}
\put(140,20){\circle*{3}}
\thinlines
\multiput(50,52.5)(0,2.5){11} {\line(1,0){90}}
\end{picture}}
\end{picture}

\begin{picture}(400,200)(0,0)
\put(0,0){\begin{picture}(200,200)(-50,-50)
\thicklines
\put(50,50){\framebox (90,30)}
\put(50,50){\line (0,-1){60}}
\put(140,50){\line (0,-1){60}}
\put(80,20){\line (1,0){60}}
\put(80,50){\line (0,-1){60}}
\put(110,20){\line (0,-1){30}}
\put(50,-20){\large 1}
\put(80,-20){\large 2}
\put(110,-20){\large 3}
\put(140,-20){\large 4}
\put(80,20){\circle*{3}}
\put(110,20){\circle*{3}}
\put(140,20){\circle*{3}}
\thinlines
\multiput(50,52.5)(0,2.5){11} {\line(1,0){90}}
\end{picture}}

\put(200,0){\begin{picture}(200,200)(-50,-50)
\thicklines
\put(50,50){\framebox (90,30)}
\put(50,50){\line (0,-1){60}}
\put(140,50){\line (0,-1){60}}
\put(50,20){\line (1,0){90}}
\put(80,50){\line (0,-1){60}}
\put(110,20){\line (0,-1){30}}
\put(50,-20){\large 1}
\put(80,-20){\large 2}
\put(110,-20){\large 3}
\put(140,-20){\large 4}
\put(50,20){\circle*{3}}
\put(110,20){\circle*{3}}
\put(140,20){\circle*{3}}
\thinlines
\multiput(50,52.5)(0,2.5){11} {\line(1,0){90}}
\end{picture}}
\put(200,-20){\Large Fig.3}
\end{picture}

 
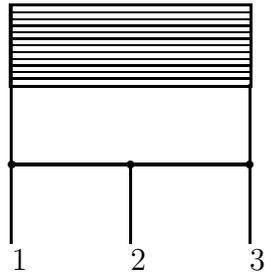
\end{document}